\documentclass[lettersize,journal]{IEEEtran}
\usepackage{amsmath,amsfonts}
\usepackage{algorithmic}
\usepackage{algorithm}
\usepackage{array}
\usepackage[caption=false,font=normalsize,labelfont=sf,textfont=sf]{subfig}
\usepackage{textcomp}
\usepackage{stfloats}
\usepackage{url}
\usepackage{verbatim}
\usepackage{graphicx}
\usepackage{cite}
\usepackage{xcolor}

\begin{document}

\title{KID Detector Readout Electronics Development for\\Habitable Worlds Observatory}

\author{Sean~Bryan$^{*1}$, Hugh Barnaby$^2$, Oketa Basha$^1$, C. Matt Bradford$^3$, Kathryn Chamberlin$^3$, Nicholas Cothard$^4$, Sumit Dahal$^4$, Thomas Essinger-Hileman$^4$, Alessandro Geist$^4$, Jason Glenn$^4$, Tracee Jamison-Hooks$^1$, Abarna Karthikeyan$^1$, Philip Mauskopf$^{1,5}$, Lynn Miles$^4$, Sanetra Bailey Newman$^4$, Cody Roberson$^1$, Karwan Rostem$^4$, Adrian Sinclair$^4$

\thanks{$^1$School of Earth and Space Exploration, Arizona State University,\\Tempe, AZ 85287 USA}
\thanks{$^2$School of Electrical, Computer, and Energy Engineering,\\Arizona State University, Tempe, AZ 85287 USA}
\thanks{$^3$NASA Jet Propulsion Laboratory, Pasadena, CA 91109 USA}
\thanks{$^4$NASA Goddard Space Flight Center, Greenbelt, MD 20771 USA}
\thanks{$^5$Department of Physics, Arizona State University, Tempe, AZ 85281 USA}
\thanks{$^*$Corresponding Author: sean.a.bryan@asu.edu}
\thanks{Manuscript received September 18th, 2025; revised MMMM DD, 2025.}}

\markboth{IEEE Transactions on Applied Superconductivity,~Vol.~36, No.~6, September~2025}%
{Shell \MakeLowercase{\textit{et al.}}: A Sample Article Using IEEEtran.cls for IEEE Journals}

%\IEEEpubid{0000--0000/00\$00.00~\copyright~2025 IEEE}

\maketitle

\begin{abstract}
We present the status and goals of the readout electronics system we are developing to support the detector arrays in the coronagraph instrument on the NASA Habitable Worlds Observatory (HWO) mission currently in development. HWO aims to revolutionize exoplanet exploration by performing direct imaging and spectroscopy of 25 or more habitable exoplanets, and to resolve a broad range of astrophysics science questions as well. Since exoplanet yield depends critically on the detector dark count rate, as we show in this paper, the ambitious goals of HWO require arrays of single-photon energy-resolving detectors. We argue that Kinetic Inductance Detectors (KIDs) are best suited to meet these requirements. To support the detectors required for HWO and future far-IR missions, at the required power consumption and detector count, we are developing a radiation-tolerant reconfigurable readout system for both imaging and energy-resolving single photon KID detector arrays. We leverage an existing RFSoC-based system we built for NASA balloons that has a power consumption of 30 Watts and reads out 2000-4000 detectors (i.e. 7-15 mW/pixel), and move to a radiation tolerant Kintex Ultrascale FPGA chip to bring low-power wide bandwidth readout to a space-qualified platform for the first time. This improves significantly over previous spaceflight systems, and delivers what is required for NASA's future needs: ~100,000 pixels with less than 1 kW total power consumption. Overall, the system we are developing is a significant step forward in capability, and retires many key risks for the Habitable Worlds Observatory mission.
\end{abstract}


\section{Introduction}
\IEEEPARstart{T}{he} Habitable Worlds Observatory (HWO)\cite{feinberg24} aims to revolutionize exoplanet exploration by performing direct imaging and spectroscopy of 25 or more habitable exoplanets, and to resolve a broad range of astrophysics science questions. The mission is bringing together the exoplanet, astrophysics, and broader astronomy community around a single shared vision of a new observatory for the 2030s. The low-risk mission concept leverages high-heritage technology from a range of previous successful space missions. One key enabling technology is high-performance detectors to image and take spectra of individual planets.

In this paper, we describe our ongoing\cite{jamison-hooks25} effort to develop radiation-hardened readout electronics to support the high-performance detectors needed for mission success in HWO. We derive requirements, and scaling of these requirements against mission and science parameters, on the dark count rate of any detector technology selected for HWO. We also trace how dark count rate affects the exoplanet yield. We show that existing MKID detectors meet the dark count requirements needed to yield 25 or more exoplanets. Finally, we describe our implementation approach.

\section{Detector Dark Counts Drive\\Exoplanet Yield in HWO}

\begin{figure*}
\begin{center}
\includegraphics[width=1.0\textwidth]{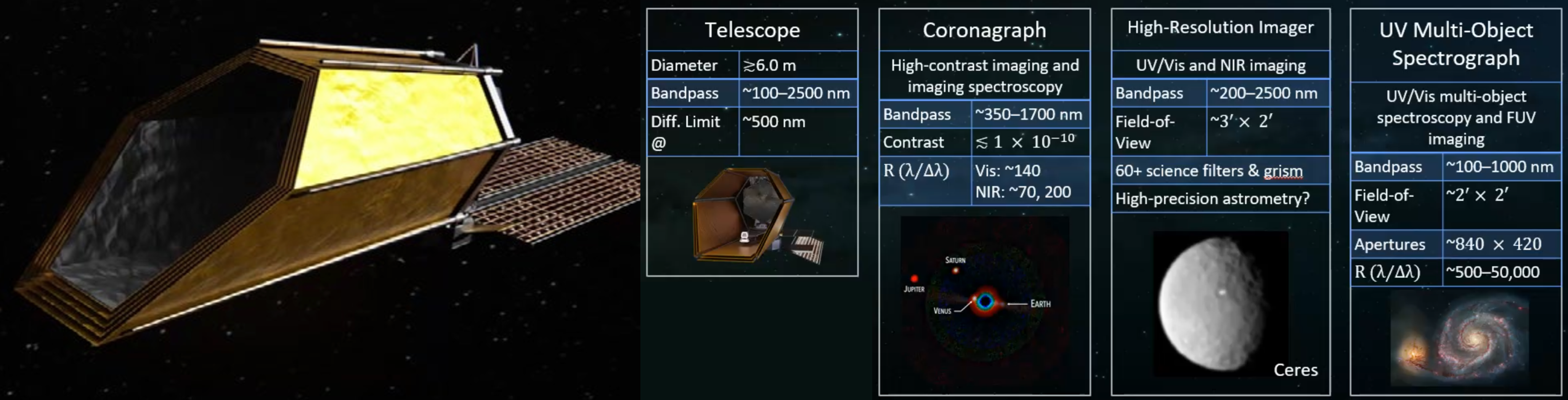}
\caption{Current conceptual rendering and key specifications of the Habitable Worlds Observatory\cite{feinberg24}.}
\label{hwo_overview}
\end{center}
\end{figure*}

As shown in the overview in Figure~\ref{hwo_overview}, the current concept for HWO has several instruments with preliminary specifications. The exoplanet direct imaging and spectroscopy science goal drives the requirements, including critical requirements for the size of the telescope and the dark count rate of the detectors as we discuss in this section. Work is ongoing \cite{mamajek23} developing a detailed target list for the mission to develop requirements. In addition, low dark counts dramatically improve\cite{steiger24,steiger25} adaptive optics and coronagraph performance compared to conventional detectors. 

Here we develop a toy model of the directly detected signal from a habitable exoplanet. This model yields similar conclusions to more complete scenario modeling\cite{howe24}, but our approach yields a single scaling equation that shows how exoplanet yield (and therefore science reach) scales with the science goals (star/planet type and range) and instrument parameters (telescope size, spectral resolution, and dark counts). This enables rapid systems engineering trade studies to support mission formulation. In our model, we estimate the photon count rate on a HWO-like detector pixel from observing an idealized earth-like planet orbiting an idealized sun-like star. We then convert this to an estimate of how the exoplanet yield scales with the telescope diameter and detector dark count rate, illustrating how critical both capabilities are for the science reach of HWO.

A photon detected by the HWO coronagraph is created as blackbody emission from the host star, reflects from the surface and atmosphere of the target planet, is collected by the HWO primary mirror, and finally detected at a pixel in the detector array. To start, the spectral flux density of an idealized blackbody host star is
\begin{equation}
B_\lambda = 4 \pi \frac{2 h c^2}{\lambda^5}\frac{1}{e^\frac{hc}{\lambda k T_{star}}-1}
\end{equation}
in units of W/(m$^2 \Delta\lambda$). Here $h,c,k$ are physical constants, $\lambda$ is the wavelength, and $T_{star}=5000$ K is taken to be the temperature of the sun-like host star. Multiplying by the surface area (taking $D_{star}=1.4 \times 10^{9}$ m to be the diameter of the sun-like star), and integrating over the bandwidth of a HWO spectral channel near the peak of the spectrum ($\lambda=$500 nm w/ 3.5 nm bandwidth), yields the total in-band power $P_{emit}$ emitted by the star. The target planet reflects a fraction of this emitted light
\begin{equation}
F_{ref} = \alpha \frac{\pi (D_{planet}/2)^2}{4 \pi d_{planet}^2}
\end{equation}
where $\alpha$ is the albedo of the planet (taken to be 0.3 to be earth-like), $D_{planet}$ is the diameter of the planet (taken to be 12,725 km to be earth-like) and $d_{planet}$ is the distance from the star to the planet (taken to be $149 \times 10^9$ m to be earth-sun-like). This reflected light then travels to the telescope a distance $d_{telescope}$. To have a survey volume large enough to yield 25 or more planets, here we take that distance to be 50 lightyears as a bounding case. The light is then collected by the primary mirror of the telescope that has a diameter $D_{telescope}$ (taken to be 6 m in line with the conceptual design). The fraction of the reflected light that is collected is therefore
\begin{equation}
F_{collected} = \frac{\pi (D_{telescope}/2)^2}{4 \pi d_{telescope}^2}.
\end{equation}
Multiplying these factors together, and converting to a photon count rate, yields
\begin{eqnarray}
\mathrm{Signal} &=& P_{emit} \times F_{ref} \times F_{collected} \times \frac{\lambda}{h c} \nonumber \\
 &=& 0.025~\mathrm{photons~per~second}.
\end{eqnarray}
This is the in-band photon rate \textit{incident} onto the telescope. The actual detected signal will be even dimmer by a factor of the telescope optical efficiency and the detector quantum efficiency.

This approximate calculation we presented above illustrates how the signal scales with key parameters. Combining terms into a single equations yields
\begin{equation}
\mathrm{Signal} \propto D_{telescope}^2  \frac{D_{planet}^2 D_{star}^2}{d_{planet}^2 d_{planet-to-telescope}^2} e^{T_{star}}.
\end{equation}
The signal scales as the square of the geometrical factors, and at the peak wavelength of the emission the signal scales approximately exponentially with star temperature. We note that within a population of candidate planetary systems, the physical properties such as the planet diameter, planet-star distance, star diameter, and star temperature all have complex interactions among themselves and are also not parameters that we can control. However, we are able to control the telescope diameter and dark count rate based on the technologies we select for the mission to deliver required exoplanet yield.

\subsection{Exoplanet Yield}
To translate this signal scaling into an exoplanet yield, we note that to efficiently measure an exoplanet the signal must be greater than the sum of the detector dark count rate and the photon rate leaking over from the inner working angle of the coronagraph. Recent developments in coronagraph technology for Roman\cite{bailey23} and other efforts are impressive, and are on track to deliver required performance for HWO soon. Assuming this progress continues, the dark count rate of the detectors themselves will become the limiting factor for the science reach of HWO.

To estimate the number of exoplanets we can observe efficiently, we calculate the the survey volume $\frac{4}{3} \pi d_{planet-to-telescope}^3$ for which the signal is greater than the dark count. The yield of planets is proportional to the survey volume. Applying this relationship gives that the number of surveyable exoplanets $N_{planets}$ scales as
\begin{equation}
N_{planets} \propto \frac{D_{telescope}^3}{(\mathrm{Dark~Counts})^{\frac{3}{2}}}  \frac{D_{planet}^3 D_{star}^3}{d_{planet}^3} e^{\frac{3}{2}T_{star}}.
\end{equation}
Quantitatively, only a modest 10\% change in each parameter multiplies out to over a factor of ten change in the exoplanet yield. This rapid scaling in the signal means that at this early stage in mission formulation relatively small uncertainties in the range and stellar properties in the target list, as well as instrument properties such as telescope diameter, will have large effects on the science reach of the mission.

This means that large margin in the detector performance, especially the dark count rate, is critical to assuring mission success in HWO.

\section{Single-Photon Energy-Resolving KID Detectors for HWO}

\begin{table*}
\centering
\caption{Key parameters for superconducting detector readout. \label{tab:Parameters}}%\vspace{-0.2in}
\begin{tabular}{l|c|l}
Parameter & Value & Driver \\
\hline\hline
Number of Tones & up to 4000 & 2 tones per resonator for UVOIR KIDs for pulse handling\\
Total Available Bandwidth & $\ge$2~GHz & Provide adequate tone spacing for large arrays\\
Minimum freq & 400~MHz & Provide 6:1 readout bandwidth far-IR tone spacing\\
Frame rate & up to 10~kHz & Pulse following and cosmic ray identification\\
Tone Tracking time & $<$2~ms  & Far-IR observation temporal bandwidth \\
Phase noise & -95 dBc/Hz & Ensures readout limited by cold amplifiers \\
ADC/DAC bit depth & $\geq$11 & Ensures bit noise remains sub-dominant\\
Per pixel power dissipation & 10 mW & System constraints of HWO, far-IR missions\\

\hline
 \end{tabular}
\end{table*}

Several detector technologies are being considered for HWO. EMCCDs\cite{khan24} are sensitive optical detectors with modest cooling requirements, but the current best demonstrated dark count rate of $\sim$0.01 photons per exposure (for exposure times 1-100 s, i.e. 0.01-1 sample per second) in the optical is not yet low enough to be subdominant to the exoplanet signal for the $\gg 1$ sample per second frame rate desired for despeckling algorithms. Linear-mode Avalanche Photodiode detectors (LmAPDs) show promise to deliver low dark counts\cite{baker23}, but are at relatively low TRL. EMCCDs have been demonstrated in balloon\cite{kyne20} flight, but not at the optical wavelengths required for HWO. LmAPDs have not yet been demonstrated in sub-orbital or spaceflight.

Single-photon-counting energy-resolving KID detectors are excellent candidates for HWO, based on their demonstrated low dark counts, broad wavelength coverage, and high TRL6 technical readiness. Detectors recently demonstrated at JPL\cite{day24} have the required dark count (4.4-6.6 $\times10^{-3}$ photons per second), wavelength coverage from optical to infrared, and a sample rate (i.e. frame rate) of 50 kilosamples per second. In several optical sub-bands, energy-resolving optical detectors have demonstrated even lower dark counts in the laboratory and deployed at a ground-based observatory\cite{swimmer23}. Kinetic Inductance detectors have flown in high-altitude balloon flights on BLAST\cite{sinclair20} and OLIMPO\cite{masi19} and are therefore TRL6.

No detector technology suitable for HWO currently has the combination of required performance and full TRL, so all detector technologies require technology maturation for HWO. The path to scale the performance of EMCCDs and LmAPDs requires improvements in fabrication and the physical processes inside the detectors themselves. Because the future scaling of these fundamental detector physics investigations is not currently known, we cannot yet assure that the dark count and frame rate capability will improve by the large factors needed to meet requirements before mission implementation. This means there is significant technical risk in baselining EMCCDs or LmAPDs for HWO at the current time.

In contrast, single-photon energy-resolving KID detectors already have the required performance, and detector readout has been demonstrated in high-altitude balloon flight. Scaling the detector technology itself towards full TRL does not require changing the physical processes in the detector itself, and is therefore low risk. Cryogenic cooling to the required temperatures has already been demonstrated in spaceflight on the Planck\cite{planck} satellite. TKID detectors\cite{florang25,rao25,wandui20} have been demonstrated in the laboratory, including devices based on the MgB$_2$ material\cite{greenfield25} with far more modest cryogenic cooling requirements. In the effort we describe in this paper, we are scaling up the detector readout to higher TRL by maturing a radiation-tolerant system. 

\section{Readout Implementation}

\begin{figure*}[b]
\centering
\begin{tabular}{r|l|l|l|}
~                          & \begin{tabular}[c]{@{}l@{}}Virtex-4QV\\ XQRV4QV\end{tabular} & \begin{tabular}[c]{@{}l@{}}Virtex-5QV\\ XQRV5QV\end{tabular} & \begin{tabular}[c]{@{}l@{}}\textbf{RT Kintex UltraScale} \\ \textbf{XQRKU060} \\ \textbf{\textit{(this work)}}\end{tabular} \\ \hline
\textit{Radiation Hardness}         & Tolerant                                                     & Hard                                                         & Tolerant                                                                              \\ \hline
\textit{Fabrication Process {[}nm{]}}           & 90                                                           & 65                                                           & \textbf{20}                                                                                    \\ \hline
\textit{Memory {[}Mb{]}}            & 4.1 to 9.9                                                   & 12.3                                                         & \textbf{38}                                                                                    \\ \hline
\textit{System Logic Cells {[}K{]}} & 55 to 200                                                    & 131                                                          & \textbf{726}                                                                                   \\ \hline
\textit{CLB Flip-Flops {[}K{]}}     & 49.1 to 178.1                                                & 81.9                                                         & \textbf{663}                                                                                   \\ \hline
\textit{CLB LUTs {[}K{]}}           & 49.1 to 178.1                                                & 81.9                                                         & \textbf{331}                                                                                   \\ \hline
\textit{Transceivers}               & None                                                         & 18 at 3.126 GB/s                                             & \textbf{32} at \textbf{12.5 Gb/s}                                                                       \\ \hline
\textit{User I/O}                   & 640 to 960                                                   & 836                                                          & \textbf{620}                                                                                   \\ \hline
\textit{DSP Slices}                 & 32 to 192                                                    & 320                                                          & \textbf{2,760}                                                                                
\end{tabular}
\caption{Based on our comparison with previous generation Virtex FPGAs (center and left columns) we selected the Kintex (right column) since it has an improved power consumption of 70\% combined with an improvement of 12x in transceiver capability and $>$5x in logic cells. Entries marked in bold indicate improved capability in our selected FPGA over previous generations.}
    \label{fpga_table}
\end{figure*}

Motivated by the need for extremely low dark counts, and the high technical readiness of KID detectors, in this program we are developing a radiation-tolerant KID detector readout for the first time. To do this, we will select a compatible FPGA and ADC/DAC, select an algorithm, and develop both the hardware and FPGA programming suitable for test and eventual infusion into future spaceflight systems.

As illustrated in Table~\ref{tab:Parameters}, our system needs a range of key performance parameters to support Habitable Worlds Observatory. To facilitate the large number of detectors in the mission, each readout supports up to 4000 detectors, including both the RF bandwidth and onboard compute needed to do so. We select ADCs and DACs with at least 11 noise-free bits to support the required at-detector readout noise level of -95 dBc/Hz. Crucially, since HWO and other future missions may require 100,000 or more detectors to meet their science requirements, our system is required to operate on 10 mW of electrical power or less per detector. This enables the readout to support the large number of required detectors in a power envelope less than 1000 W. To do this, we scale up the performance of the BLAST\cite{gordon16} readout system that already meets this per-detector power requirement.

\subsection{Selected FPGA and ADC/DAC}

We selected new AMD/Xilinx RFSoC Kintex Ultrascale XQRKU060 radiation tolerant FPGA chip. As shown in Figure~\ref{fpga_table} AMD/Xilinx Ultrascale FPGAs are designed for ultrahigh processing throughput with low power and offer advantages over previous generations of chips. Our overall architecture is similar to the existing RFSoC system, but will include external radiation hard analog components required for space operations.

\subsection{Algorithm}

\begin{figure*}
\begin{center}
\includegraphics[width=1.0\textwidth]{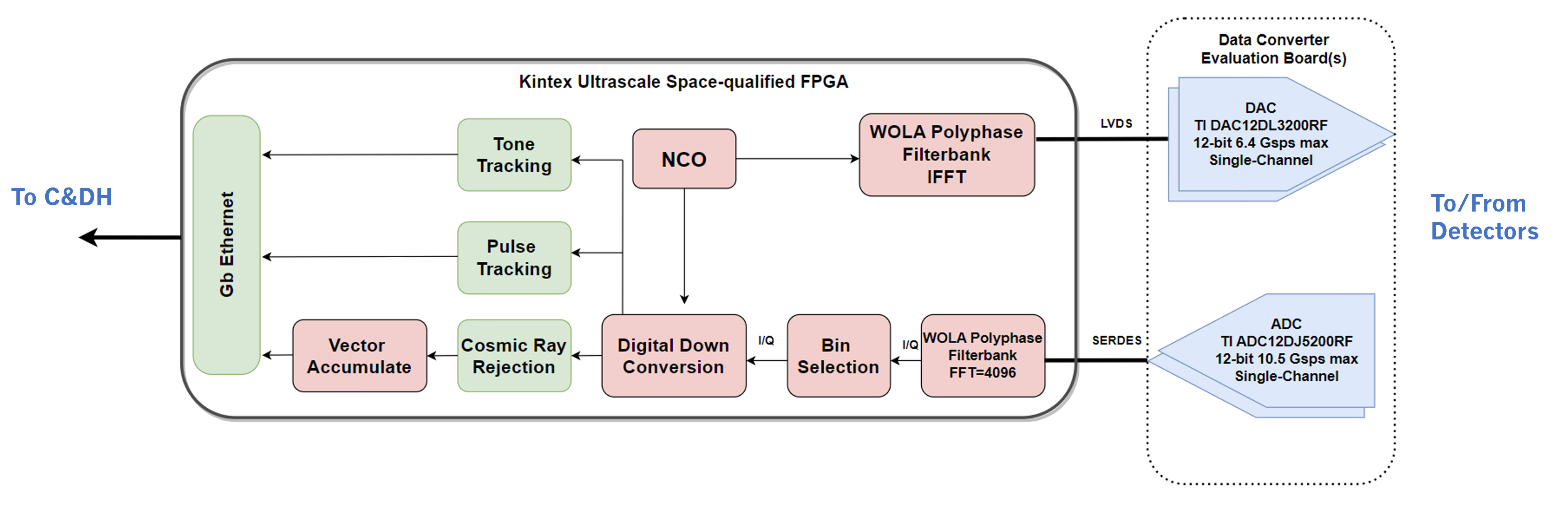}
\caption{Block diagram of our algorithm }
\label{block_diagram}
\end{center}
\end{figure*}

At a high level, Figure ~\ref{block_diagram} shows a flowchart of our system to and from the detectors, and interfacing with the C\&DH computer for eventual downlink to the ground. Our approach is similar to proven approaches used in BLAST (and described in more detail in the BLAST detector readout paper\cite{gordon16}) and other missions, but we have made changes to support future mission needs that we describe here. We are using a Numerically-Controlled Oscillator (NCO) and an IFFT to generate the probe tones for each detector onboard, instead of the approach for BLAST where we had a fixed precomputed set of tone signals. This lets us retune the detectors in realtime using tone tracking if we choose, as shown in the block diagram. Not shown in the block diagram, this also facilitates rapid VNA-style frequency sweeps that let us detect and characterize the detectors upon system bootup and during subsequent retunings. 

After the tones pass through the analog front end and the detectors, and are sampled by the ADC, we then FFT the data. Both when synthesizing tones with the IFFT, and here at the FFT, we use a polyphase filterbank to limit the far frequency sidelobes of each FFT channel. We re-use the coefficients for both the FFT and IFFT polyphase filterbanks. We describe our polyphase filterbank elsewhere\cite{basha25} in these proceedings. We then use a bin selection and digital downconversion step that are similar to the BLAST firmware. For digital downconversion, we re-use the signals from the NCO that were used to synthesize the tones at the start of the algorithm.

After digital down conversion, we currently baseline that each individual detector will be sampled at 10 kHz. The detectors\cite{day24} we currently baseline were sampled in the laboratory at 50 kHz and delivered outstanding dark counts and optical performance. Our readout is reconfigurable for higher sample rates if required. Next in our algorithm, we detect pulses in the signals using a baseline-removal filter, then a matched filter, and a threshold trigger. This has a reconfigurable purpose. For single photon counting detector readout, this is the endpoint of the signal chain. In that configuration, we detect, characterize, and calculate a timestamp for each pulse, and that information is sent to the gigabit ethernet and ultimately downlinked to the ground as the main science data. In the imaging detector configuration however, we infill the data during a detected pulse with a moving-average, then proceed to the vector accumulate block to average and downsample the data before we send the entire timestream to the gigabit ethernet for downlink as the main science data.

Also optionally, the signal is sent from the digital downconversion block to a tone tracking block. This calculates a running average power level on each detector, and optionally retunes each detector to a frequency closer to the optimal point for the current average power level. Elsewhere in these proceedings, a group\cite{rowe25} is investigating whether this retuning should be continuous or in discrete steps to deliver optimal performance, but our readout is compatible with both approaches.

\subsection{Flight Hardware}

\begin{figure*}
\begin{center}
\begin{tabular}{cc}
\includegraphics[height=0.3\textwidth, clip=true, trim=0in 0in 0in 0in]{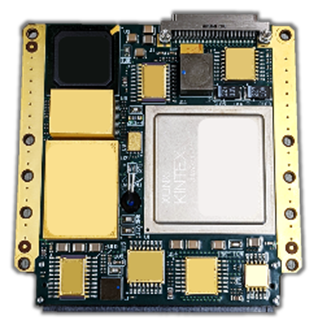} &
\includegraphics[height=0.3\textwidth, clip=true, trim=0in 0in 0in 0in]{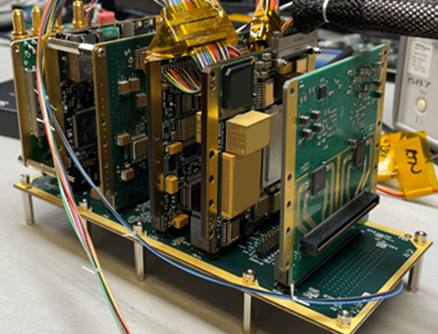}
\end{tabular}
\caption{The proposed readout system builds upon high-TRL modular SpaceCube hardware.  \textit{Left panel:} The readout system uses a build-to-print SpaceCube v3.0 Mini FPGA card (primary side pictured) housing a Xilinx Kintex KU060 chip. \textit{Right panel:} Photograph of a full SpaceCube electronics system with 1U CubeSat cards inserted into a backplane. }
\label{spacecube}
\end{center}
\end{figure*}

We build on the SpaceCube hardware shown in Figure~\ref{spacecube} and the ongoing effort led at NASA Goddard to raise the TRL of key electronics components by flying them on cubesats attached to the standard SpaceCube bus. In this program, we have already tested an earlier generation FPGA chip. Our goal is to design our readout to be compatible with this bus to enable future flight test and further TRL maturation activities.

\section{Conclusion}
Direct imaging and spectroscopy of earth-like exoplanets with Habitable Worlds Observatory will revolutionize the field. This ambitious science goal requires detector arrays with low dark counts, and MKID detectors are in an excellent position to meet these requirements. Our radiation-hardened MKID detector readout system leverages heritage from other missions, and we are on track to implement and demonstrate the system in a laboratory and beamline environment.

\section*{Acknowledgments}
The research described in this paper was carried out under a contract with the National Aeronautics and Space Administration (80NSSC25K7184).

We acknowledge input from Benjamin Mazin.

\bibliographystyle{IEEEtran}
\bibliography{report} 

@article{swimmer23,
author = {Noah Swimmer and W. Hawkins Clay and Nicholas Zobrist and Benjamin A. Mazin},
journal = {Opt. Express},
number = {6},
pages = {10775--10793},
publisher = {Optica Publishing Group},
title = {Characterizing the dark count rate of a large-format MKID array},
volume = {31},
month = {Mar},
year = {2023},
url = {https://opg.optica.org/oe/abstract.cfm?URI=oe-31-6-10775},
doi = {10.1364/OE.485003},
}

@inproceedings{khan24,
author = {Aafaque R. Khan and Erika Hamden and Gillian Kyne and April D. Jewell and John Henessey and Shouleh Nikzad and Vincent Picouet and Olivia Jones and Harrison Bradley and Nazende Kerkeser and Zeren Lin and Brock Parker and Grant West and John Ford and Frank Gacon and Dave Beaty and Jacob Vider},
title = {{Advancing ultraviolet detector technology for future missions: investigating the dark current plateau in silicon detectors using photon-counting EMCCDs}},
volume = {13093},
booktitle = {Space Telescopes and Instrumentation 2024: Ultraviolet to Gamma Ray},
editor = {Jan-Willem A. den Herder and Shouleh Nikzad and Kazuhiro Nakazawa},
organization = {International Society for Optics and Photonics},
publisher = {SPIE},
pages = {130930O},
year = {2024},
doi = {10.1117/12.3021039},
URL = {https://doi.org/10.1117/12.3021039}
}

@inproceedings{sinclair20,
author = {A. Sinclair and S. B. Gordon and P. Ade and P. Ashton and H.-M. Cho and G. Coppi and A. Corso and M. Devlin and S. Dicker and B. Dober and L. Fissel and N. Galitzky and J. Gao and C. Groppi and G. C. Hilton and J. Hubmayr and K. Irwin and J. Klein and D. Li and N. Lourie and I. Lowe and H. Mani and P. D. Mauskopf and C. McKenney and F. Nati and G. Novak and E. Pascale and G. Pisano and J. Soler and C. Tucker},
title = {{In-flight performance of the BLAST-TNG Kinetic Inductance Detector arrays and Readout Electronics}},
booktitle = {Proceedings of the 31st Symposium on Space Terahertz Technology},
organization = {ISSTT},
pages = {16},
year = {2020},
}

@inproceedings{jamison-hooks25,
author = {T. Jamison-Hooks and L. Miles and S. Newman-Bailey and O. Basha and A. Karthikeyan and S. E. Kay and S. Bryan and P. Mauskopf and T. Essinger-Hileman and J. Glenn and S. Dahal and A. Sinclair and K. Chamberlain},
title = {{Development of Space Qualified Signal Processing Readout Electronics for HabWorlds and Origins Space Telescope Detector and Arrays}},
booktitle = {Proceedings of the 34th Symposium on Space Terahertz Technology},
organization = {ISSTT},
year = {2025},
URL = {doi.org/10.48550/arXiv.2508.00322}
}

@article{day24,
  title = {A 25-micrometer Single-Photon-Sensitive Kinetic Inductance Detector},
  author = {Day, Peter K. and Cothard, Nicholas F. and Albert, Christopher and Foote, Logan and Kane, Elijah and Eom, Byeong H. and Basu Thakur, Ritoban and Janssen, Reinier M. J. and Beyer, Andrew and Echternach, Pierre M. and van Berkel, Sven and Hailey-Dunsheath, Steven and Stevenson, Thomas R. and Dabironezare, Shahab and Baselmans, Jochem J. A. and Glenn, Jason and Bradford, C. Matt and Leduc, Henry G.},
  journal = {Phys. Rev. X},
  volume = {14},
  issue = {4},
  pages = {041005},
  numpages = {20},
  year = {2024},
  month = {Oct},
  publisher = {American Physical Society},
  doi = {10.1103/PhysRevX.14.041005},
  url = {https://link.aps.org/doi/10.1103/PhysRevX.14.041005}
}

@article{planck,
	author = {{Tauber, J. A.} and {Nielsen, P. H.} and {Martín-Polegre, A.} and {Crill, B.} and {Cuttaia, F.} and {Ganga, K.} and {Gudmundsson, J.} and {Jones, W.} and {Lawrence, C.} and {Meinhold, P.} and {Norgaard-Nielsen, H. U.} and {Oxborrow, C. A.} and {Partridge, B.} and {Roudier, G.} and {Sandri, M.} and {Scott, D.} and {Terenzi, L.} and {Villa, F.} and {Bernard, J. P.} and {Burigana, C.} and {Franceschi, E.} and {Kurki-Suonio, H.} and {Mandolesi, N.} and {Puget, J. L.} and {Toffolatti, L.}},
	title = {Characterization of the in-flight properties of the Planck telescope},
	DOI= "10.1051/0004-6361/201833150",
	url= "https://doi.org/10.1051/0004-6361/201833150",
	journal = {Astronomy and Astrophysics},
	year = 2019,
	volume = 622,
	pages = "A55",
}

@article{masi19,
doi = {10.1088/1475-7516/2019/07/003},
url = {https://dx.doi.org/10.1088/1475-7516/2019/07/003},
year = {2019},
month = {jul},
publisher = {},
volume = {2019},
number = {07},
pages = {003},
author = {Masi, S. and de Bernardis, P. and Paiella, A. and Piacentini, F. and Lamagna, L. and Coppolecchia, A. and Ade, P.A.R. and Battistelli, E.S. and Castellano, M.G. and Colantoni, I. and Columbro, F. and D'Alessandro, G. and Petris, M. De and Gordon, S. and Magneville, C. and Mauskopf, P. and Pettinari, G. and Pisano, G. and Polenta, G. and Presta, G. and Tommasi, E. and Tucker, C. and Vdovin, V. and Volpe, A. and Yvon, D.},
title = {Kinetic Inductance Detectors  for the OLIMPO experiment:  in-flight operation and performance},
journal = {Journal of Cosmology and Astroparticle Physics}
}

@inproceedings{baker23,
author = {Ian Baker and Matthew Hicks and Chris Maxey and Daniel Owton},
title = {{Leonardo UK high performance shortwave APDs for astronomy}},
volume = {12687},
booktitle = {Infrared Sensors, Devices, and Applications XIII},
editor = {Priyalal Wijewarnasuriya and Arvind I. D'Souza and Ashok K. Sood},
organization = {International Society for Optics and Photonics},
publisher = {SPIE},
pages = {1268702},
year = {2023},
doi = {10.1117/12.2679478},
URL = {https://doi.org/10.1117/12.2679478}
}

@inproceedings{florang25,
author = {I. Fogarty Florang and others},
title = {{Thermal kinetic inductance detector physics: understanding devices 
through measurements of position dependence and first-principles models}},
booktitle = {These Proceedings},
year = {2025}
}

@inproceedings{steiger25,
author = {S. Steiger and others},
title = {{Wavefront Sensing and Control with Superconducting Detectors for the Habitable Worlds Observatory}},
booktitle = {These Proceedings},
year = {2025}
}

@inproceedings{steiger24,
author = {Sarah Steiger and Laurent Pueyo and Emiel H. Por and Pin Chen and R{\'e}mi Soummer and Rapha{\"e}l Pourcelot and Iva Laginja and Vanessa P. Bailey},
title = {{Simulated performance of energy-resolving detectors towards exoplanet imaging with the Habitable Worlds Observatory}},
volume = {13092},
booktitle = {Space Telescopes and Instrumentation 2024: Optical, Infrared, and Millimeter Wave},
editor = {Laura E. Coyle and Shuji Matsuura and Marshall D. Perrin},
organization = {International Society for Optics and Photonics},
publisher = {SPIE},
pages = {130921W},
year = {2024},
doi = {10.1117/12.3020603},
URL = {https://doi.org/10.1117/12.3020603}
}

@article{howe24,
author = {Alex R. Howe and Christopher C. Stark and John E. Sadleir},
title = {{Scientific impact of a noiseless energy-resolving detector for a future exoplanet-imaging mission}},
volume = {10},
journal = {Journal of Astronomical Telescopes, Instruments, and Systems},
number = {2},
publisher = {SPIE},
pages = {025008},
year = {2024},
doi = {10.1117/1.JATIS.10.2.025008},
URL = {https://doi.org/10.1117/1.JATIS.10.2.025008}
}

@inproceedings{bailey23,
author = {Vanessa P. Bailey and Eduardo Bendek and Brian Monacelli and Caleb Baker and Gasia Bedrosian and Eric Cady and Ewan S. Douglas and Tyler Groff and Sergi R. Hildebrandt and N. Jeremy Kasdin and John Krist and Bruce Macintosh and Bertrand Mennesson and Patrick Morrissey and Ilya Poberezhskiy and Hari B. Subedi and Jason Rhodes and Aki Roberge and Marie Ygouf and Robert T. Zellem and Feng Zhao and Neil T. Zimmerman},
title = {{Nancy Grace Roman Space Telescope coronagraph instrument overview and status}},
volume = {12680},
booktitle = {Techniques and Instrumentation for Detection of Exoplanets XI},
editor = {Garreth J. Ruane},
organization = {International Society for Optics and Photonics},
publisher = {SPIE},
pages = {126800T},
year = {2023},
doi = {10.1117/12.2679036},
URL = {https://doi.org/10.1117/12.2679036}
}

@inproceedings{rao25,
author = {T. Rao and others},
title = {{Progress developing thermal kinetic inductance detectors for charged particle detection in neutron beta decay experiments }},
booktitle = {These Proceedings},
year = {2025}
}

@inproceedings{rowe25,
author = {S. Rowe and others},
title = {{Mitigating intermodulation distortion in multi-tone resonator readouts via frequency rounding}},
booktitle = {These Proceedings},
year = {2025}
}

@inproceedings{basha25,
author = {O. Basha and others},
title = {{
A Design Methodology for an FPGA-based Polyphase Filterbank for Microwave Kinetic Inductance Detector (MKID) Readout Systems }},
booktitle = {These Proceedings},
year = {2025}
}

@inproceedings{mamajek23,
author = {E. Mamajek and K. Stapelfeldt},
title = {{NASA Exoplanet Exploration Program (ExEP) Mission Star List for the Habitable Worlds Observatory}},
booktitle = {doi.org/10.48550/arXiv.2402.12414},
year = {2023}
}

@article{kyne20,
author = {Gillian Kyne and Erika T. Hamden and Shouleh Nikzad and Keri Hoadley and April D. Jewell and Todd J. Jones and Michael E. Hoenk and Samuel R. Cheng and D. Christopher Martin and Nicole R. Lingner and David Schiminovich and Bruno Milliard and Robert Grange and Olivier Daigle},
title = {{Delta-doped electron-multiplying CCDs for FIREBall-2}},
volume = {6},
journal = {Journal of Astronomical Telescopes, Instruments, and Systems},
number = {1},
publisher = {SPIE},
pages = {011007},
year = {2020},
doi = {10.1117/1.JATIS.6.1.011007},
URL = {https://doi.org/10.1117/1.JATIS.6.1.011007}
}

@article{wandui20,
    author = {Wandui, Albert and Bock, James J. and Frez, Clifford and Hollister, M. and Minutolo, Lorenzo and Nguyen, Hien and Steinbach, Bryan and Turner, Anthony and Zmuidzinas, Jonas and O’Brient, Roger},
    title = {Thermal kinetic inductance detectors for millimeter-wave detection},
    journal = {Journal of Applied Physics},
    volume = {128},
    number = {4},
    pages = {044508},
    year = {2020},
    month = {07},
    issn = {0021-8979},
    doi = {10.1063/5.0002413},
    url = {https://doi.org/10.1063/5.0002413},
    eprint = {https://pubs.aip.org/aip/jap/article-pdf/doi/10.1063/5.0002413/15252752/044508\_1\_online.pdf},
}

@article{greenfield25,
    author = {Greenfield, J. and Bell, C. and Faramarzi, F. and Kim, C. and Basu Thakur, R. and Wandui, A. and Frez, C. and Mauskopf, P. and Cunnane, D.},
    title = {Kinetic inductance and non-linearity of MgB2 films at 4K},
    journal = {Applied Physics Letters},
    volume = {126},
    number = {2},
    pages = {022602},
    year = {2025},
    month = {01},
    issn = {0003-6951},
    doi = {10.1063/5.0245866},
    url = {https://doi.org/10.1063/5.0245866},
    eprint = {https://pubs.aip.org/aip/apl/article-pdf/doi/10.1063/5.0245866/20351998/022602\_1\_5.0245866.pdf},
}

@article{gordon16,
author = {Gordon, Samuel and Dober, Brad and Sinclair, Adrian and Rowe, Samuel and Bryan, Sean and Mauskopf, Philip and Austermann, Jason and Devlin, Mark and Dicker, Simon and Gao, Jiansong and Hilton, Gene C. and Hubmayr, Johannes and Jones, Glenn and Klein, Jeffrey and Lourie, Nathan P. and McKenney, Christopher and Nati, Federico and Soler, Juan D. and Strader, Matthew and Vissers, Michael},
title = {An Open Source, FPGA-Based LeKID Readout for BLAST-TNG: Pre-Flight Results},
journal = {Journal of Astronomical Instrumentation},
volume = {05},
number = {04},
pages = {1641003},
year = {2016},
doi = {10.1142/S2251171716410038}
}

@inproceedings{feinberg24,
author = {Lee Feinberg and John Ziemer and Megan Ansdell and Julie Crooke and Courtney Dressing and Bertrand Mennesson and John O'Meara and Joshua Pepper and Aki Roberge},
title = {{The Habitable Worlds Observatory engineering view: status, plans, and opportunities}},
volume = {13092},
booktitle = {Space Telescopes and Instrumentation 2024: Optical, Infrared, and Millimeter Wave},
editor = {Laura E. Coyle and Shuji Matsuura and Marshall D. Perrin},
organization = {International Society for Optics and Photonics},
publisher = {SPIE},
pages = {130921N},
year = {2024},
doi = {10.1117/12.3018328},
URL = {https://doi.org/10.1117/12.3018328}
}

\newpage

\vfill

\end{document}